\documentclass[a4paper,11pt]{article}
\usepackage{jheppub}
\usepackage{natbib}
\usepackage{graphicx}
\usepackage{bm}
\usepackage{amssymb}
\usepackage{color}
\usepackage{amsmath}
\usepackage{mathrsfs} 
\usepackage{amstext}
\usepackage[english]{babel}
\usepackage{latexsym}
\usepackage{array}             
\usepackage{multirow}         
\usepackage[usenames,dvipsnames]{xcolor}

\newcommand{\beq}{\begin{equation}}
\newcommand{\eeq}{\end{equation}}
\newcommand{\beqa}{\begin{eqnarray}}
\newcommand{\eeqa}{\end{eqnarray}}

\begin{document}

\title{Quantum analog to flapping of flags: interface instability for co-flow binary superfluids}

\author[a,b]{Yu-Ping An}%
\author[a,b,c,d]{Li Li}
 \affiliation[a]{CAS Key Laboratory of Theoretical Physics, Institute of Theoretical Physics,
Chinese Academy of Sciences, Beijing 100190, China}
\affiliation[b]{School of Physical Sciences, University of Chinese Academy of Sciences, Beijing 100049, China}
\affiliation[c]{School of Fundamental Physics and Mathematical Sciences, Hangzhou Institute for Advanced Study, University of Chinese Academy of Sciences, Hangzhou 310024, China}
\affiliation[d]{Peng Huanwu Collaborative Center for Research and Education, Beihang University, Beijing 100191, China}
 \author[e,f]{Hua-Bi Zeng}
 \affiliation[e]{Center for Theoretical Physics , Hainan University, Haikou 570228, China}
 \affiliation[f]{  Center for Gravitation and Cosmology, College of Physical Science and Technology, Yangzhou University, Yangzhou 225009, China}
\emailAdd{anyuping@itp.ac.cn, liliphy@itp.ac.cn, zenghuabi@hainanu.edu.cn}
 \vspace{1cm}

\abstract{We study the interface dynamics in immiscible binary superfluids using its holographic description, which naturally consists of an inviscid superfluid component and a viscous normal fluid component. We give the first theoretical realization of interface instability for two superfluid components moving with identical velocity, providing a quantum analog to the flapping of flags that is common in daily life. This behavior is in sharp contrast to the one from Gross-Pitaevskii equation for which no such co-flow instability develops in an isolated uniform system because of Galilean invariance. The real time evolution triggered by the dynamical instability exhibits intricate nonlinear patterns leading to quantum turbulence reminiscent of the quantum Kelvin-Helmholtz instability. Moreover, we show that such interface dynamics is essentially different from the Landau instability for which the frictionless flow becomes thermodynamically unstable above a critical superfluid velocity. Our study uncovers the rich interface dynamics of quantum fluids and the emergence of complex flow phenomena.}

\maketitle

\section{Introduction}
Interface instabilities are ubiquitous in nature where they are found across various length scales and material properties. Examples range from the formation of ocean waves, to the generation of swirling patterns on Jupiter, to the trigger mechanism for pulsar glitches of neutron stars, and to the vortex formation in atomic Bose-Einstein condensates (BECs). Efforts to rationalize their occurrence in nature date back hundreds of years, see~\cite{Zahn2013,Drazin_2002,chandrasekhar2013hydrodynamic} for comprehensive introduction. Thus far, whether there exist robust mechanisms underpinning the genesis of such instabilities is still an open question. Of particular interest is interface dynamics in quantum fluids, for which novel  interface dynamics and patterns could develop due to the quantum characteristics of quantum fluids. Describing and understanding the behavior of interface dynamics for quantum fluids far from equilibrium, going therefore beyond the hydrodynamic approximation, is a question of extreme importance with applications ranging from condensed matter physics to astronomy. 

Inspiration from natural phenomena potentially gives rise to a variety of unexpected phenomena in quantum case. A common everyday life example of classical interface instabilities is the flapping of
flags in wind--the instability of the passive deformable membrane between two distinct parallel streams with the same density and the same velocity. This dynamic instability was unravelled by Rayleigh~\cite{Rayleigh}, see a recent experiment in~\cite{RN466}. This stimulates interest in studying analogical phenomenon in quantum fluids, for which superfluids are the proper ideal objects where this idea can be implemented without reservations. In the quantum fluid case, the role of the flag is played by the interface of immiscible binary superfluids, while the parallel streams correspond to the moving binary superfluids.

There have been numerous theoretical efforts on the interface instabilities addressing the dynamics of binary superfluids in the literature. However, most of these studies rely on the mean-field Gross-Pitaevskii equation (GPE), which is a model equation applicable in weakly interacting regime. Moreover, it is challenging for considering finite temperature effects and dissipation. On the other hand, holography provides a natural tool to incorporate these aspects, representing strongly coupled quantum many-body systems at finite temperature and dissipation through gravitational systems of black holes with one higher dimension. Holographic superfluids have been extensively studied, including phenomena like superfluid turbulence~\cite{Adams:2012pj,Du:2014lwa}, dark solitons~\cite{Guo:2018mip}, and the Kibble-Zurek mechanism~\cite{Chesler:2014gya,Sonner:2014tca,delCampo:2021rak}. Nevertheless, there are few investigations considering the interface dynamics in superfluids by holography. 
The counterflow instability in immiscible binary superfluids was studied in~\cite{An:2024ebg} recently using numerical holography, where the wave number of the fastest growing modes yields a non-monotonic dependence of the relative velocity, in sharp contrast to the one from GPE. 

To study the quantum analogy of flapping of flags, we consider the interface of binary superfluids where the two superfluid components move with the same velocity relative to their normal fluid components. While this co-flow instability was suggested to develop in superfuilds using a heuristic argument, no theoretical realization has been given in the literature. In particular, we show that this instability can not be presented from GPE. In this work, we provide the first concrete realization of such co-flow interface instability in strongly coupled immiscible binary superfluids at finite temperature and chemical potential using its holographic description. We further show that the interface dynamics has nothing to do with the Landau instability that associated with the excitation of negative energies~\cite{landau}. These results are directly testable in laboratory experiments.

\section{Holographic setup}
We begin by setting up the gravity description of a two-component superfluid in two spatial dimensions~\cite{An:2024ebg}. The action is formulated as
    \begin{equation}\label{model}
        \begin{aligned}
            S=\int dx^4\sqrt{-g}\left[\frac{1}{2\kappa_N^2}(R+\frac{6}{L^2})+\mathcal{L}_m\right],
          \end{aligned}  
    \end{equation}
with 
    \begin{equation}
        \begin{aligned}
             \mathcal{L}_m=&-(\mathcal{D}_\mu\Psi_1)^*
            \mathcal{D}^\mu\Psi_1-m_1^2|\Psi_1|^2-(\mathcal{D}_\mu\Psi_2)^*
            \mathcal{D}^\mu\Psi_2-m_2^2|\Psi_2|^2\\&-\frac{\nu}{2}|\Psi_1|^2|\Psi_2|^2-\frac{1}{4}F^{\mu\nu}F_{\mu\nu}\,,
          \end{aligned}  
    \end{equation}
where $\mathcal{D}_\mu\Psi_i=(\nabla_\mu-ie_iA_\mu)\Psi_i$, $A_\mu$ is the $U(1)$ gauge field with $F_{\mu\nu}$ its strength. The two bulk scalars $\Psi_1$ and $\Psi_2$ are charged under the same $U(1)$ gauge field and correspond to the two superfluid components of the dual system. The inter-component coupling $\nu$ characterizes the interaction between two components of superfluids and determines the miscibility of the superfluids~\cite{An:2024ebg}. To consider the interface dynamics, we focus on $\nu>0$, \emph{i.e.} immiscible binary superfluids. Early studies on holographic binary orders can be found, \emph{e.g.} in~\cite{Basu:2010fa,Cai:2013wma,Yang:2019ibe,Yao:2022fov}.

At finite temperature, each of the two liquids contains both superfluid and normal components. This appears to be a rather complicated problem. To avoid these complications, we consider the case where the fluctuations
of the temperature and the normal fluid’s velocity are frozen. This corresponds to the probe limit in the gravity description for which the back-reaction of matter content to the geometry is neglected. More precisely,  background is given by the 
Schwarzschild AdS black brane.
    \begin{equation}\label{backg}
        ds^2=\frac{L^2}{z^2}(-f(z)dt^2-2dtdz+dx^2+dy^2)\,,
    \end{equation}
with the blackening function $f(z)=1-(z/z_h)^3$ vanishing at the event horizon $z=z_h$.
This corresponds to a heat bath with temperature $T=3/(4\pi z_h)$ on the boundary. Note that in the probe limit, the fluctuations of the temperature and the normal fluid’s velocity are frozen. In practice, we set $L=z_h=1$ and choose the radial gauge $A_z=0$. We initially employ the holographic superfluid with two identical components by choosing $m_1^2=m_2^2=-2$, $e_1=e_2=1$, corresponding to the dual scalar operators that has the scaling dimension $\Delta=2$.

After solving the bulk equations of motion (see Appendix~\ref{app:eoms}), we can read off all relevant observables by the standard holographic dictionary. The gravity dual thus provides a first-principle description of superfluid dynamics. In particular, near the AdS boundary $z=0$, the asymptotic expansions can be presented as
    \begin{equation}\label{UV}
        \begin{aligned}
        A_\mu&=a_\mu-b_\mu z+\mathcal{O}(z^2),\quad
        \Psi_i=\Psi_i^{(v)} z^2+\mathcal{O}(z^3),\quad i=1,2\,,
        \end{aligned}
    \end{equation}
where we have already turned off the leading source term of each scalar field so as to realize the superfluid state for which the $U(1)$ symmetry is broken spontaneously. Therefore, $\Psi_i^{(v)}$ corresponds to the superfluid condensate $\mathcal{O}_i $. Moreover, $a_t=\mu$ is the chemical potential and $b_t=\rho$ is the charge density. $\bm{a}=(a_x,a_y)$ are related to the superfluid velocity $\bm{v}_i^s\equiv(v_{ix}^s, v_{iy}^s)=\nabla\theta_i-\bm{a}$ with $\theta_i$ the phase of the superfluid condensation $\mathcal{O}_i$. For the sake of brevity, we use bold-face letters to denote vectors in boundary spatial directions. We will set $a_x=a_y=0$, such that the superfluid velocity is given by $\bm{v}_i=\nabla\theta_i$ for each superfluid component. 

Due to the scaling symmetry of the dynamics
\begin{equation}
(t,x,y,z)\rightarrow \lambda(t,x,y,z), \quad z_h\rightarrow \lambda z_h,\quad (T,\mu,\bm{v}_i^s)\rightarrow \frac{1}{\lambda}(T,\mu, \bm{v}_i^s),\quad (\rho,\mathcal{O}_i)\rightarrow \frac{1}{\lambda^2}(\rho,\mathcal{O}_i)\,,
\end{equation}
$T$ and $\mu$ are not independent quantities. Since we have fixed $z_h=1$ (\emph{i.e.}, $T=3/4\pi$), $\mu$ is the only free parameter for given theory parameters. There is a second order phase transition when $\mu\ge \mu_c\simeq 4.064$ by choosing $z_h=1$. This also fixes the ratio $T/T_c=\mu_c/\mu$.  
The system involves two complex scalar operators $\mathcal{O}_i$ which carry the same charge under a global $U(1)$ symmetry. Below a critical temperature $T_c$, the scalar operators develop a nonzero expectation value spontaneously breaking the $U(1)$ symmetry and driving the system into a superfluid phase.

\begin{figure}[htpb]
        \centering
            \includegraphics[width=0.49\linewidth]{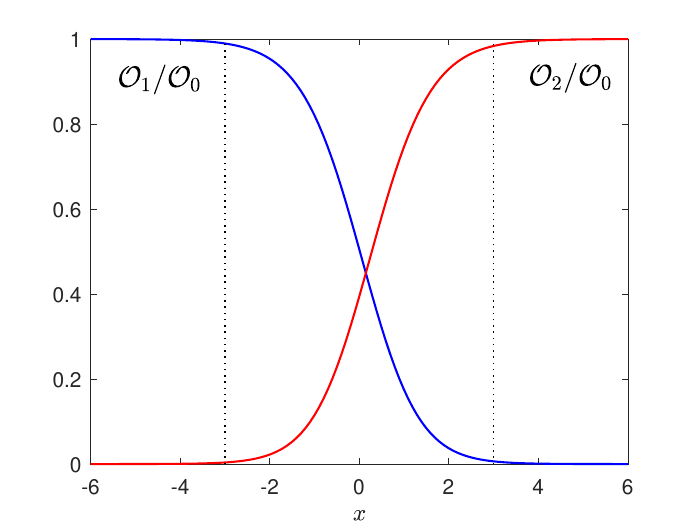}
            \includegraphics[width=0.49\linewidth]{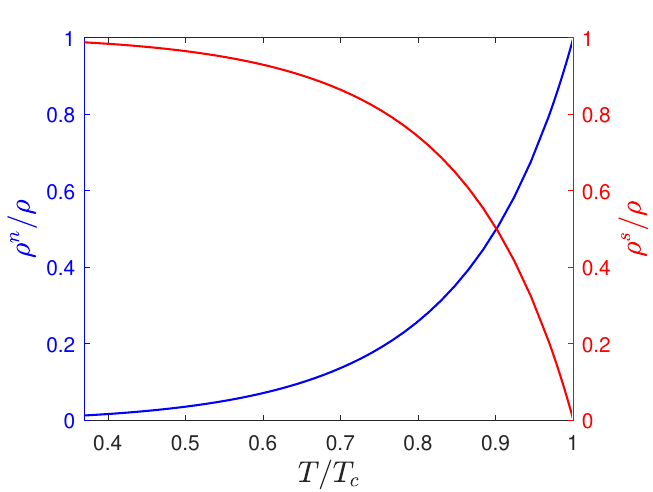}
            \caption{Stationary configuration for the immiscible binary superfluids. \textbf{Left:} The normalized order parameter of the two superfluid components $\mathcal{O}_i$ for $v_y=5.348T_c$ with $\mathcal{O}_0$ the value of the order parameter
            far from the interface. The phase with a interface consists of the configuration in which one component occupies half of
            the space and the other does the rest. \textbf{Right:} The fraction $\rho^s/\rho$ carried by the superfluid condensate and the one $\rho^n/\rho$ by the normal fluid component as a function of temperature, where $\rho=\rho^s+\rho^n$. The superfluid component density and the one for the normal fluid component are read from the region far from the interface for which only one of the superfluid condensations develops spontaneously below $T_c$.}
    \label{O_rho}
\end{figure}

\section{Stationary phase-separated configuration}
For immiscible binary superfluids, the two superfluid components form a stationary phase-separated configuration. Without loss of generality, we consider the case where they undergo phase separation and form an straight interface at $x=0$ together with a superfluid velocity $v_y$ along the $y$ axis. 

To obtain the stationary phase-separated configuration, we choose the following ansatz around the bulk geometry~\eqref{backg}.
\begin{equation}
\label{ansatzps}
\begin{split}
&\Psi_i=z\phi_i(z,x)e^{i\Theta_i(z,x,y)}, \quad A_t=A_t(z,x)\,,\\
&V_{iy}(z,x)\equiv\partial_y\Theta_i-A_y,\quad\quad  V_{ix}(z,x)\equiv\partial_x\Theta_i-A_x=0\,,
\end{split}
\end{equation}
with the gauge $\partial_z\Theta_i=-{A_t}/{f}$ for the stationary state. Note that the phase $\theta_i$ of the condensation $\mathcal{O}_i$ corresponds to $\Theta_i|_{z=0}$. Since the superfluid velocity is irrotational and $v_{ix}^s=0$, we have $\partial_x v_{iy}^s=\partial_y v_{ix}^s=0$. Therefore, $v_{iy}^s$ should be independent of $x$ on the boundary. According to the holographic dictionary, we have $V_{1y}|_{z=0}=V_{2y}|_{z=0}=v_y$, where $v_y$ is the superfluld velocity relative to the thermal bath. 

The resulted bulk equations of motion are given by
\begin{equation}
\begin{split}
    \partial_z(f\partial_z\phi_i)+\partial_x^2\phi_i-(v_i)_y^2\phi_i+\frac{A_t^2}{f}\phi_i-z\phi_i-\frac{\nu}{2}\phi_j^2\phi_i=0\,,\\
        f\partial_z^2A_t+\partial_x^2A_t-2A_t(\sum_i\phi_i^2)=0\,,\\
\partial_z(f\partial_zV_{iy})+\partial_x^2V_{iy}-2\sum_i V_{iy}\,\phi_i^2=0\,,      
\end{split}
\end{equation}
which involves five coupled PDEs for $(\phi_i, V_{iy}, A_t)$ that all depend on the variables $z$ and $x$. 
We solve the above equations numerically by the Newton-Raphson method. In the radial direction, we adopt the Chebyshev pseudo-spectral method with the following boundary conditions at $z=0$:
\begin{equation}
    \begin{aligned}
        \phi_i(z=0)=0,\quad A_t(z=0)=\mu\,,\\ \quad V_{1y}(z=0)=V_{2y}(z=0)=v_y\,,
    \end{aligned}
\end{equation}
as well as the regular condition at $z=1$. In the $x$ direction, we choose the fourth order finite difference scheme and the Neumann boundary condition far from the interface. 

The stationary phase-separated configuration for which both the suprflows have the same velocity moving relative to the normal component is shown in the right panel of Fig.~\ref{O_rho}. A sharp interface is manifest due to the strong competition between the two superfluid condensations. Moreover, far away from the interface, only one of the superfluid condensations exists. The superfluid component density can be read from holographic dictionary as $\rho^s=\lim_{v^s\rightarrow 0}\mu b_y/v^s$~\cite{PhysRevD.82.026001}, supposing the superfluid velocity is along $y$ direction. Then the normal component density is $\rho^n=\rho-\rho^s$. The fraction $\rho^s/\rho$ and $\rho^n/\rho$ as a function of temperature are presented in the right panel of Fig.~\ref{O_rho}. It is reminiscent of the temperature dependence of the superfluid and normal components of liquid He II as measured from in the torsional oscillation disk stack experiment.

\section{Interface dynamics from GPE}
\label{sec:GPE}
In this section we show the main results from the two-component GEPs for phase-separated binary BECs for comparison. In the mean field approximation, the condensate wave functions are described by $\Psi_j(x,y,t)=\sqrt{n_j(x,y,t)}e^{i\theta_j}(\mathbf{r},t)$ with $n_j$ and $\theta_j$ the particle density and phase of the $j$-th component of the binary superfluids. The coupled GPEs without an external potential are given as~\cite{PhysRev.134.A543}:
    \begin{equation}
    \label{GP_2}
        i\partial_t \Psi_i=(-\frac{1}{2m_i}\nabla^2-\mu_i+g_i|\Psi_i|^2+g_{ij}|\Psi_j|^2)\Psi_i,\quad (i,j=1,2, \quad i\ne j)\,.
    \end{equation}
Here $m_j$ is the atomic mass and $g_1, g_2$ and $g_{12}$ are the coupling
constants. We consider $g_{12}>\sqrt{g_1 g_2}$ for immiscible BECs and the ground state for the binary condensates with an equal particle number. 

The configuration with a stationary phase-separated interface takes
    \begin{equation}
    \label{ansatz_GP}
    \Psi_i(x,y)=\psi_i(x)e^{im_iv_i^s y}\,,
    \end{equation}
with $v_i^s$ the superfluid velocity of the $i$-th superfluid component along the $y$ direction. Substitute~\eqref{ansatz_GP} into~\eqref{GP_2}, one gets the following time-independent GPEs for $\psi_i(x)$:
\begin{equation}
    (-\frac{1}{2m_i}\partial_x^2-\mu_i+\frac{m_iv_i^{s2}}{2}+g_i|\psi_i|^2+g_{ij}|\psi_j|^2)\psi_i=0,\quad (i,j=1,2, \quad i\ne j)\,.
\end{equation}
Far from the interface, one has $\partial_x\psi_i=0$, $\psi_j=0$ ($i\neq j$) and therefore $\psi_i=(\mu_i-m_iv_i^{s2}/2)/g_i=\sqrt{n_i}$ with $\mu_i$ the chemical potential. Profiles for $\psi_i(x)$ solved from
these equations are similar to those shown in the left panel of Fig.~\ref{O_rho}.

Once the stationary solutions are given, one can study the interface stability using the linear perturbation analysis. We turn on the  perturbations around the
stationary background $\psi_i^0$, \emph{i.e.},
\begin{equation}
\label{wavefunction}
    \Psi_i=\left[\psi_i^0(x)+u_i(x)e^{iky-i\omega t}-v_i^*(x)e^{-iky+i\omega^* t}\right]e^{im_iv_i^s y}\,,
\end{equation}
where $\omega$ is the frequency and $k$ is the wave number of the excitation along the translationally invariant $y$ axis. Then we obtain the BdG eqution by linearizing the time-dependent GPEs~\eqref{GP_2}:
\begin{equation}\label{Hamton}
    \mathcal{H}\mathbf{U}=\omega\mathbf{U}\,,
\end{equation}
\begin{equation}
    \mathcal{H}=\left\{
\begin{tabular}{cccc}
$h_1^+$&$-g_1(\psi_1^0)^2$&$g_{12}\psi_1^0\psi_2^0$&$-g_{12}\psi_1^0\psi_2^0$\\
$g_1(\psi_1^0)^2$&$-h_1^-$&$g_{12}\psi_1^0\psi_2^0$&$-g_{12}\psi_1^0\psi_2^0$\\
$g_{12}\psi_1^0\psi_2^0$&$-g_{12}\psi_1^0\psi_2^0$&$h_2^+$&$-g_2(\psi_2^0)^2$\\
$g_{12}\psi_1^0\psi_2^0$&$-g_{12}\psi_1^0\psi_2^0$&$g_2(\psi_2^0)^2$&$-h_2^-$
\end{tabular}
\right\},
\end{equation}
with $\mathbf{U}=(u_1,w_1,u_2,w_2)^\mathrm{T}$ and
\begin{equation}
    h_i^\pm=-\frac{1}{2m_i}\left[\partial_x^2-(k\pm m_iv_i^s)^2\right]-\mu_i+2g_i|\psi_i^0|^2+g_{ij}|\psi_j^0|^2.
\end{equation}
By numerically diagonalizing this discretized BdG Hamiltonian $\mathcal{H}$, one can get the eigenfrequency $\omega$ for each wave number $k$. Since this Hamiltonian $\mathcal{H}$ is real, one has $\mathcal{H}\mathbf{U}^*=\omega^*\mathbf{U}^*$, \emph{i.e.}, $\omega^*$ is also an eigenvalue once $\omega$ is an eigenvalue. Therefore, the system becomes dynamically unstable whenever the frequency takes a non-vanishing imaginary part $\mathrm{Im}(\omega)\ne 0$.

\begin{figure}[htpb]
        \centering
            \includegraphics[width=0.7\linewidth]{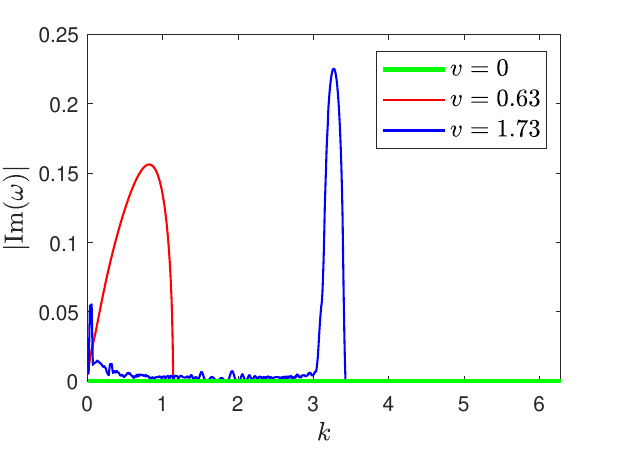}
            \caption{$|\mathrm{Im}(\omega)|$ versus $k$ for superflow velocity $v_1^s=-v_2^s=v$ obtained from the GPEs. The system becomes dynamically unstable whenever $\mathrm{Im}(\omega)\neq 0$. Green, red and blue lines correspond to $v=0$, $0.63$ and $1.73$, respectively. It is manifest that there is no instability for $v=0$. We have chosen $g=m=1$, $g_{12}=2$ and $\mu=0.5$.}
    \label{v_k_GP}
\end{figure}

Before considering the case with $v_1^s=v_2^s$, let's review the case with $v_1^s=-v_2^s=v$ first. It corresponds to an straight interface at $x=0$ together with a relative velocity $2v$ along the $y$ axis.
By calculating $\mathrm{Im}(\omega)$ for different $v$ and $k$, one can extract the wave number $k_0$ of the most unstable mode, 
which can be compared directly with those of our holographic theory. We limit ourselves to the parameters as $g_1=g_2=g$, $m_1=m_2=m$, and $\mu_1-m_1v_1^{s2}/2=\mu_2-m_2v_2^{s2}/2=\mu$. 
The results are shown in Fig.~\ref{v_k_GP} for three velocities $v=0$, $0.63$ and $1.73$. Note that there is no instability for $v=0$.
For each $v>0$, there is a leading unstable mode $k_0$ which has the largest $|\mathrm{Im}(\omega)|$. For small velocity $k_0\sim v^2$ that corresponds to the Kelvin-Helmholtz instability, and for large velocity $k_0\sim v$ dominated by the counter-flow instability~\cite{crossover}. These results stand in sharp contrast to the holographic model, which gives a non-monotonic relation between $k_0$ and $v$~\cite{An:2024ebg}.

Besides above counter-flow instability
in an isolated uniform system with $v_1^s=-v_2^s=v$, a particularly interesting case is the interface dynamics without relative velocity, \emph{i.e.}, $v_1^s=v_2^s=v$. However, since GPEs of~\eqref{GP_2} are Galilean covariant, such case is equivalent to that with zero relative velocity after a coordinate transformation. To be specific, the transformed wave function is
\begin{equation}
    \Psi_i'(y',t')=e^{-imvy+i\frac{mv^2}{2}t}\Psi_i(y,t)=\psi_ie^{i\frac{mv^2}{2}t' },
\end{equation}
with $y'=y-vt$ and $t'=t$. One can show directly that the BdG equation~\eqref{Hamton} for perturbations of $\Psi'$ in transformed coordinates is the same as that of $\Psi$ with zero $v$, but with $\mu'=\mu-mv^2/2$. 
Then one finds that the eigenfrequency $\omega$ from~\eqref{Hamton} for the transformed wave function always has a vanishing imaginary part, as is shown in Fig.~\ref{v_k_GP}, and therefore the phase-separated interface in this case is dynamical stable. Some instabilities can be triggered by introducing an external potential or an impurity by hand, which can essentially break the Galilean boost invariance, see \emph{e.g.}~\cite{PhysRevA.70.013608,PhysRevA.66.013610,PhysRevA.62.061601,PhysRevLett.100.160402,PhysRevA.80.053602,PhysRevA.79.053619}. Nevertheless, they are different from the co-flow instability we show in next section.

Note that for a single-component superfluid at zero temperature, a superfluid can flow at any velocity without creating excitations due to boost invariance in an infinite space, assuming there is no external potential or any impurity. As we have just shown above, the boost symmetry of the GEPs~\eqref{GP_2} also removes the interface instability for the binary superfluids without relative velocity. Nevertheless, by heating the system, a superfluid can move with a velocity relative to its normal component. This destroys the boost invariance of the superfluids. 
The Landau criterion suggests that there should exist a critical velocity beyond which excitations will be spontaneously generated everywhere and the system should develop a genuine instability. For an infinite system, the crucial point from Landau’s argument above is the presence of the normal fluid component in a superfluid, which is definitely absent from GPEs~\eqref{GP_2} as it takes account for the superfluid component only. Therefore, one may wonder if the interface instability can develop for the case without relative velocity after considering the normal component at finite temperature. As we will show in next section, this co-flow instability does occur.
\begin{figure}[htpb]
        \centering
            \includegraphics[width=0.7\linewidth]{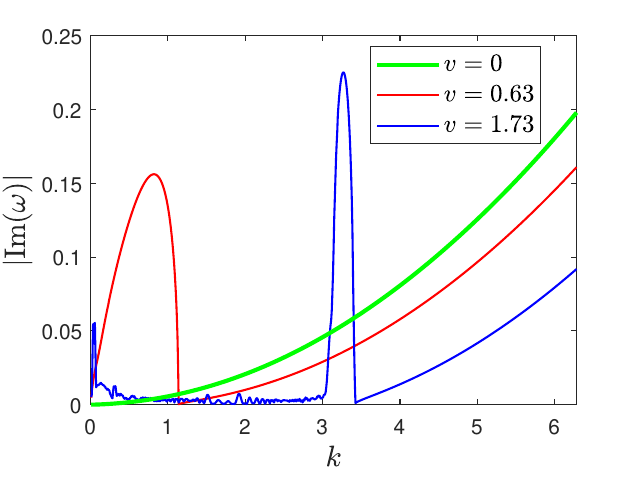}
            \caption{The same quantities are shown as in Fig.~\ref{v_k_GP} with the same parameters, except that additional dissipation $\gamma=0.01$ is included. This gives nonphysical result for co-flow interface instability: instability strength grows fast and unboundedly as $k$ increases.}
    \label{v_k_GP_dissipation}
\end{figure}

We also note that dissipation or finite temperature effect in GPE is often considered by adding a phenomenological dissipation term in~\eqref{GP_2}:
\begin{equation}
     i\partial_t \Psi_i\rightarrow i\partial_t \Psi_i-\gamma \partial_t \Psi_i\, ,
\end{equation}
with $\gamma$ a constant. The effect of this additional parameter $\gamma$ of~\eqref{Hamton} is simply $\omega\rightarrow \omega'=(1-i\gamma)\omega$. This leads to nonphysical effect of co-flow instability. That is, at large momentum, instability strength increases unboundedly as $k$ increases, see Fig.~\ref{v_k_GP_dissipation}. Compared to this naive phenomenological model, the holographic model gives more natural and physical result.

\section{Interface instability for co-flow binary superfluids}
We now consider the interface instability for two superfluid components moving with identical velocity in our holographic setup. The conventional phenomenological Landau-Tisza model considers the coexistence of a superfluid component with a normal fluid one. Above a critical temperature the superfluid component vanishes, while at $T=0$ the normal component vanishes yielding a system with a pure superfluid. At finite temperature, a superfluid and a normal fluid component coexist. The interface of the binary superfluids could be unstable at finite temperature without relative velocity due to the ``wind" from the relative velocity between the superfluid component $\rho^s$ and normal fluid component $\rho^n$. In our holographic model, the black hole background corresponds to a heat bath on the boundary, so finite temperature effect is naturally incorporated, see the right panel of Fig.~\ref{O_rho} for the
fraction $\rho^s/\rho$ and $\rho^n/\rho$ as a function of temperature. We shall uncover the dynamical interface instability for binary superfluids moving with the same velocity $v_y$, providing a quantum analog to the flapping of flags in wind.

\begin{figure*}[htpb]
        \centering
            \includegraphics[width=0.99\linewidth]{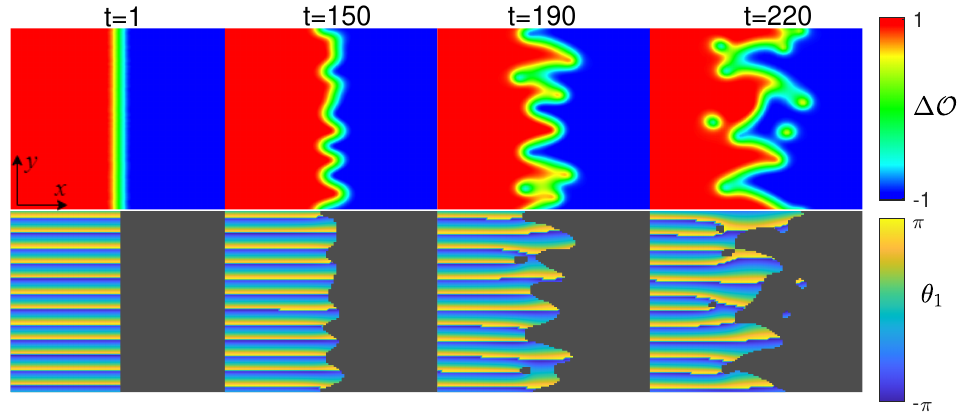}
            \caption{Interface dynamics for two superfluid components moving with the same velocity $v_y=5.348T_c$ at $T/T_c=0.677$. \textbf{Upper panel}: Snapshots of the condensation difference $\Delta\mathcal{O}=(|\mathcal{O}_1|^2-|\mathcal{O}_2|^2)/|\mathcal{O}_0|^2$ at different times. \textbf{Bottom panel}: The corresponding snapshots of the phase $\theta_1$ of the first superfluid component. Small initial perturbations on the interface destabilize and evolve into larger amplitude structures, leading to the formation of vortices and the onset of quantum turbulence.  We only show $\theta_1$ for $|\mathcal{O}_1|^2-|\mathcal{O}_2|^2>0$ since otherwise $|\mathcal{O}_1|$ is small and $\theta_1$ is pure noise. We choose $\nu=1$ and maintain $\mu=6$. Plotted region is $[-20,20]\times[0,40]$.}
    \label{fancy}
\end{figure*}

We start with the initial nondissipative state in Fig.~\ref{O_rho}, corresponding to the thermal equilibrium in the presence of the interface and superflows. In thermal equilibrium, the normal component is at rest in the heat bath, $\bm{v}_{1}^n=\bm{v}_{2}^n=0$, while the superfluids move along the interface with velocities $\bm{v}_{1}^s=\bm{v}_{2}^s=(0, v_y)$. We turn on a small random noise to the initial stationary condensates to initiate the dynamical instability. The system size $(L_x,L_y)$ is large enough such that the boundary effect is negligible for the physics we are concerned with here. Numerical details can be found in Appendix~\ref{app:eoms}. Fig.~\ref{fancy} presents a representative example for the interface dynamics. The two superfluids share an initially flat boundary and flow with the same speed $v_y$ parallel to the interface. The instability gradually develops causing the boundary to become wavy. The amplitude of the sinusoidal interface wave consistently increases, driven by the exponential growth of initial perturbations. This situation is very similar to the phenomenon of a flapping flag in wind. Nevertheless, in the subsequent nonlinear evolution, the interface eventually undergoes disintegration into bubble-like domains of the condensates due to the quantum characteristics of quantum fluids. The phase profile $\theta_1$ of the first component reveals that each bubble-like domain contains a quantized vortex, as evident in the bottom panel at $t=220$. Similar dynamics are observed across different temperatures and relative velocities. The observed patterns are very reminiscent of quantum Kelvin-Helmholtz instability.
\begin{figure}[htpb]
    \centering
        \includegraphics[width=0.6\linewidth]{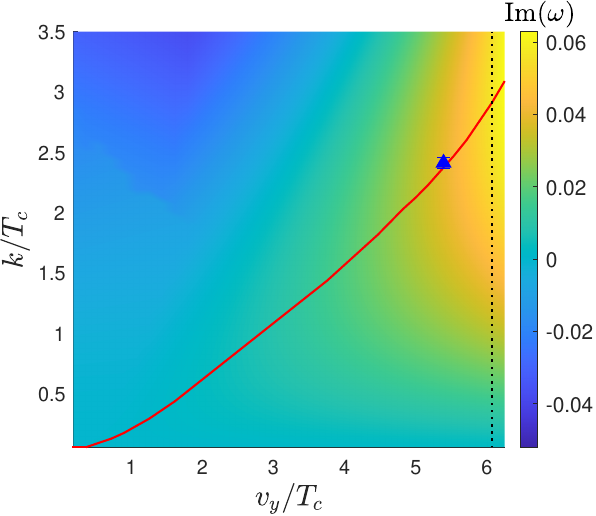}
        \caption{Illustration for the dynamical interface instability via QNMs around the stationary phase-sperated configuration with $T/T_c=0.677$ and $\nu=1$. The density plot shows $\mathrm{Im}(\omega)$ in terms of $k$ and $v_y$ with $T/T_c=0.677$ for which the fastest growing modes are denoted by the red curve. The blue triangle corresponds to $k_0$ extracted from time evolution at the same velocity as in Fig.~\ref{fancy}. The black vertical dashed line denotes the critical velocity $v/T_c=6.06$ beyond which the Landau instability begins to appear. }
    \label{k-vT}
\end{figure}

The onset of instability can be revealed through linear response analysis around the phase-separated configurations depicted in Fig.~\ref{O_rho}. Mapping to the dual gravitational description, we compute the quasi-normal modes (QNMs) on top of the stationary solutions. Thanks to the time translation symmetry and the translation symmetry along the $y$ direction, we turn on small perturbations that are decomposed in terms of Fourier modes $e^{-i(\omega t-ky)}$ with $\omega$ and $k$ the frequency and wave number of the interface wave, respectively. This results in a generalized eigenvalue problem that can be solved numerically, see Appendix~\ref{app:QNM} for more details. The quasi-normal frequency $\omega$ generically takes a complex value due to dissipation of the system. For each $k$, one can find a discrete spectrum of QNMs, for which the dominant mode has the largest value of the imaginary part Im($\omega)$. The background is dynamically unstable once Im($\omega)$ is positive. The larger the positive imaginary part is, the more unstable the system becomes. Thus, the wave number of the fastest-growing mode corresponds to a positive Im($\omega$) which reaches a maximum at a certain wave number $k=k_0$.

We demonstrate the spectrum of QMNs as a function of $k$ and $v_y$ in Fig.~\ref{k-vT} where the fastest-growing modes are denoted by the red curve. The linear analysis agrees quantitatively with the one from fully dynamical evolution, as illustrated by the blue triangle extracting from real time evolution of Fig.~\ref{fancy}. We anticipate that the co-flow interface instability would be rooted in the relative motion between the superfluid and normal fluid components. Therefore, this instability would disappear at $T=0$ for which the normal component vanishes. Indeed, we show that for GPE that has no normal fluid component, no such co-flow instability develops in an isolated uniform system because of Galilean invariance. Moreover, adding a phenomenological dissipation term to GPE results in spurious modes. Since the fraction of normal fluid component decreases quickly at low temperatures (see the right panel of Fig.~\ref{O_rho}), $k_0$ and the overall instability would decrease as the system is cooled down. These features are confirmed from our numerical computation. 

\begin{figure*}[htpb]
        \centering
            \includegraphics[width=0.99\linewidth]{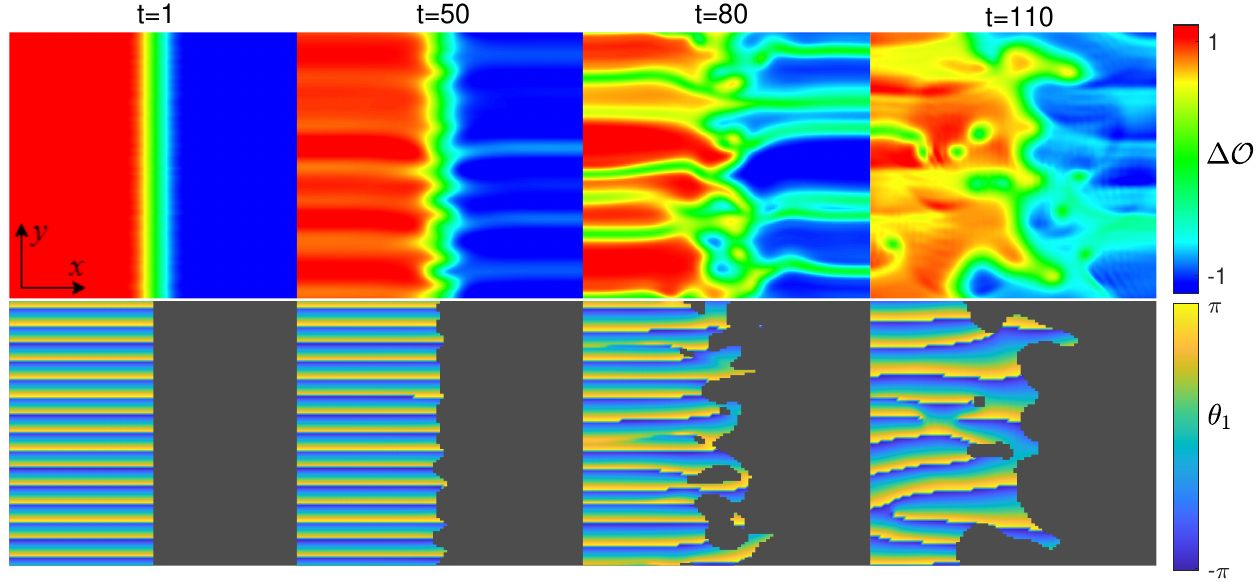}
            \caption{Interface patterns for two superfluid components moving beyond the Landau critical velocity.  \textbf{Upper panel}: Snapshots of the condensation difference $\Delta\mathcal{O}=(|\mathcal{O}_1|^2-|\mathcal{O}_2|^2)/|\mathcal{O}_0|^2$ at different times. \textbf{Bottom panel}: The corresponding dynamics for the   phase $\theta_1$ of the first superfluid component. The excitations develop everywhere and grow quickly, destabilizing not only the interface but also the whole system. 
            We choose the same parameters and plotted region as Fig.~\ref{fancy}, except for the value of the superfluid velocity $v_y/T_c=6.685$.}\label{LI}
\end{figure*}

Generally, when the superfluid component moving relative to the normal component above a critical velocity $v_c^L(T)$, the Landau instability appears (see \emph{e.g.}~\cite{Amado:2013aea,2020arXiv201006232L,Gouteraux:2022qix,2023arXiv231208243A} for early studies in holographic superfluids). The interface instability shown in Fig.~\ref{fancy} (blue triangle of Fig.~\ref{k-vT}) is well below $v_c^L$ (vertical dashed line in Fig.~\ref{k-vT}). Moreover, the excitations due to the Landau instability spontaneously generated everywhere, while the instability of Fig.~\ref{fancy} is localized around the interface. Thus, the interface dynamics we show is not from the Landau instability. We also find that the interface instability does show up in the stable region of the second sound, where the second sound remains real without a positive imaginary part. In our present setup, the criterion for thermodynamic instability yields $\partial_{v_y}[(\rho^s_1+\rho^s_2)v_y]|_{v_y=v_c}=0$ at a critical velocity $v_c$. Far from the interface, the system in thermal equilibrium is described by the homogeneous phase with a single condensate, for which the above criterion gives precisely the Landau critical velocity $v_c^L$. In contrast, for the immiscible superfluids, it does not give the critical velocity for the onset of the interface instability. Please refer to Appendix~\ref{thermo} for more details.
Therefore, we believe that the instability presented in Fig.~\ref{fancy} should be due to the intrinsic interface dynamics rather than a thermal-instability-driven one.

We illustrate the nonlinear time evolution for the superflow above the Landau critical velocity $v_c^L$ in Fig.~\ref{LI}, which shows a drastically different pattern from the one below $v_c^L$ (see Fig.~\ref{fancy}). For $v_y>v_c^L$, interface instability is accompanied with the Landau instability. This thermal instability results in the generation of spatially random excitations, and therefore, in the subsequent nonlinear evolution, the whole configuration becomes highly non-equilibrium and inhomogeneous everywhere, as evident in the upper panel at $t=110$ of Fig.~\ref{LI}. In contrast, for $v_y<v_c^L$ the dynamics develops near the interface and the regions far from the interface still stay in equilibrium as shown in Fig.~\ref{fancy}. Moreover, comparing with Fig.~\ref{fancy}, the excitations of Fig.~\ref{LI} develop much more quickly due to the coexistence of the Landau and interface instabilities.

\section{Conclusion and discussion}We have investigated the interface dynamics of strong coupling two-component superfluids flowing in the same direction with identical velocity. In contrast to the GPE that takes account of superfluid component only, our holographic model naturally incorporates both an inviscid superfluid component and a viscous normal fluid component. We have shown that due to relative velocity with respect to the normal fluid component, the interface between the two superfluid phases becomes dynamically unstable, exhibiting patterns reminiscent of quantum Kelvin-Helmholtz instability (see Fig.~\ref{fancy}). This is the first explicit realization of such co-flow interface instability in quantum fluids.

We have also demonstrated that such interface instability is not thermal instability like Landau instability for which a superfluid exchanges momentum with a ``rigid" environment. In contrast, our interface dynamics is purely an ``internal" instability of the isolated system of two superfluids without any influence from the external environment. These findings could be tested in experimental platforms such as strongly coupled ultracold Bose gases or thin helium films at low temperatures.  For example, the immiscible binary Bose-Einstein condensates
have been already achieved in a $^{85}$Rb-$^{87}$Rb system~\cite{bcsexperiment1} and a $^{87}$Rb-$^{133}$Cs system~\cite{bcsexperiment2}. Meanwhile, it is desirable to understand the mechanism behind the co-flow interface instability we find. While we have shown that this instability will be suppressed as the temperature is cooled down, we find that the surface tension $\sigma$ of the interface between
two liquids plays an opposite role. More precisely, the instability is suppressed as $\sigma$ is increased. Nevertheless, $\sigma$ itself depends on $T$ and increases by decreasing $T$. It is important to go beyond the the probe approximation used in this work. This will allow one to study the interface dynamics at zero temperature limit for which we anticipate no co-flow instability. This requires to take into account the back-reaction of the matter fields onto the metric and to use numerical relativity to determine the evolution of the system. 

Our study broadens the application of holography to non-equilibrium phenomena with finite temperature and dissipation, opening avenues for investigating interface instabilities in holography laboratories. This provides an intriguing platform to uncover the complex and unique characteristics of quantum fluid instabilities and their rich phenomena. There are several avenues for further research, extending our study and shedding light on the intricate behaviors of interface dynamics. For example, the vortex formation by surface instability is a rather generic phenomenon. We find that quantum vortices formed at the interface in the present case is much sparser compared to the case where the two superfluid components have opposite velocities~\cite{An:2024ebg}. The precise mechanism of the vortex formation is still unclear and our results highlight the importance of understanding the vortex formation and their effect on the interface dynamics. It would be very interesting to study the dependence of the instability on parameters such as rotation and magnetic field.

\acknowledgments
We thank Wei-Can Yang and Tao Shi for valuable 
discussions.
This work is partly supported by the National Natural Science Foundation of China No.12122513, No.12075298 and No.12275233. We acknowledge the use of the High Performance Cluster at Institute of Theoretical Physics, Chinese Academy of Sciences.

\appendix 

\section{Equations of motion and numerical details}\label{app:eoms}
The equations of motion for $\Psi_i$ and $A_\mu$ are given as
 \begin{equation}
    \mathcal{D}_\mu \mathcal{D}^\mu\Psi_i-m_i^2\Psi_i-\frac{\nu}{2}|\Psi_j|^2\Psi_i=0, \quad(i,j=1,2\quad i\ne j),
 \end{equation}
 \begin{equation}
     \nabla_\mu F^{\mu\nu}=-2\mathrm{Im}(\sum_i\Psi^*_i\mathcal{D}^\nu\Psi_i).
 \end{equation}
with $\mathrm{Im}$ representing imaginary part. 
Then, the explicit form of equations of motion on the top of~\eqref{backg} reads
    \begin{equation}
        \begin{aligned}
            \label{phi}
            2\partial_t\partial_z\Phi_i-[2i A_t\partial_z\Phi_i+i \partial_zA_t\Phi_i+\partial_z(f\partial_z\Phi_i)-z\Phi_i
            +\partial_x^2\Phi_i+\partial_y^2\Phi_i
            -i (\partial_xA_x+\partial_yA_y)\Phi_i&\\
            -(A_x^2+A_y^2)\Phi_i-2i (A_x\partial_x\Phi_i+A_y\partial_y\Phi_i)
            -\frac{\nu}{2}|\Phi_j|^2\Phi_i]=0, \qquad(i,j=1,2\quad i\ne j)&
        \end{aligned}
    \end{equation}
    \begin{equation}
        \label{Ax}
        \begin{aligned}
            2\partial_t\partial_zA_x-[\partial_z(\partial_xA_t+f\partial_zA_x)+\partial_y(\partial_yA_x-\partial_xA_y)-2A_x\sum_i|\Phi_i|^2
            +2\mathrm{Im}(\sum_i\Phi_i^*\partial_x\Phi_i)]=0,
        \end{aligned}
    \end{equation}
    \begin{equation}
        \label{Ay}
        \begin{aligned}           
        2\partial_t\partial_zA_y-[\partial_z(\partial_yA_t+f\partial_zA_y)+\partial_x(\partial_xA_y-\partial_yA_x)-2A_y\sum_i|\Phi_i|^2
            +2\mathrm{Im}(\sum_i\Phi_i^*\partial_y\Phi_i)]=0,
        \end{aligned}
    \end{equation}
    \begin{equation}
        \label{At}
        \begin{aligned}
            \partial_t\partial_zA_t-[\partial_x^2A_t+\partial_y^2A_t+f\partial_z(\partial_xA_x+\partial_yA_y)-\partial_t(\partial_xA_x+\partial_yA_y)
            -2A_t\sum_i|\Phi_i|^2&\\
            -2f\mathrm{Im}(\sum_i\Phi_i^*\partial_z\Phi_i)+2\mathrm{Im}(\sum_i\Phi_i^*\partial_t\Phi_i)]=0,&
        \end{aligned}
    \end{equation}
    \begin{equation}
        \label{constraint}
        \begin{aligned}     
        \partial_z(\partial_xA_x+\partial_yA_y-\partial_zA_t)-2\mathrm{Im}(\sum_i\Phi_i^*\partial_z\Phi_i)=0,
        \end{aligned}
    \end{equation}
where we have introduced $\Phi_i=\Psi_i/z$. One can find that
    \begin{equation}
    \label{relation}
        -\partial_t\mathrm{Eq.}(\ref{constraint})-\partial_z\mathrm{Eq.}(\ref{At})+\partial_x\mathrm{Eq.}(\ref{Ax})+\partial_y\mathrm{Eq.}(\ref{Ay})=2\mathrm{Im}(\sum_i\mathrm{Eq.}(\ref{phi})\times\Phi_{0i}^*)\,,
    \end{equation}
and thus the above equations are not independent.

To observe the interface instability of two-component superfluids, we perturb the stationary configurations $\Phi_{0i}$ shown \emph{e.g.} in Fig.~\ref{O_rho}. Since different initial conditions lead to quantitatively similar late-time patterns, we focus on initial condition of perturbed $\Phi_i$ that takes the following form:
\begin{equation}
    \Phi_i=\Phi_{0i}[1+\alpha\sum_k \mathrm{exp}(iky+i\theta_k)]\,,
\end{equation}
where $\theta_k$ is a random phase for each $k$, and $\alpha$ is a small number which we set to be 0.01. 
The system is then evolved with this initial configuration for $\Phi_i$ while keeping $\delta A_x=\delta A_y=0$.

Our numerical scheme is implemented as following. We adopt the fourth order Runge-Kutta method along the time direction. First, we use \eqref{phi}, \eqref{Ax} and \eqref{Ay} to evolve $\Phi$, $A_x$ and $A_y$ with the boundary condition
\begin{equation}
 \Phi(z=0)=A_x(z=0)=A_y(z=0)=0\,.   
\end{equation}
Then we use \eqref{At} to evolve $\partial_zA_t$ on the boundary. More precisely, \eqref{At} at the AdS boundary $z=0$ gives
\begin{equation}
[\partial_t\partial_zA_t-\partial_z(\partial_xA_x+\partial_yA_y)]|_{z=0}=0\,. 
\end{equation}
Note that $-\partial_zA_t(z=0)$ is nothing but the charge density or number density $\rho$ of the dual field theory.
Finally, we use \eqref{constraint} to solve $A_t$ by evolved $\Phi$, $A_x$, $A_y$ together with the boundary condition
\begin{equation}
\partial_zA_t(z=0)=-\rho,\quad A_t(z=0)=\mu\,.   
\end{equation}
Thanks to the coordinate system we use, no boundary condition is imposed at the event horizon $z_h$. 

We also have to fix the boundary conditions in other two spatial directions. In the $y$ direction we use the period boundary condition with its size $L_y$, \emph{i.e.}
\begin{equation}
\Phi(y)=\Phi(y+L_y), A_\mu(y)=A_\mu(y+L_y)\,.
\end{equation}
We impose the Neumann boundary condition 
\begin{equation}
\partial_x\Phi(x=\pm L_x/2)=\partial_x A_\mu(x=\pm L_x/2)=0\,,
\end{equation}
along the $x$ direction, with $L_x$ the size of the system. The size of our system $L_x\times L_y$ is prepared properly for each parameter set such that 
that the boundary effect can be negligible for the physics we are interested in.

To discretize the coupled PDEs~\eqref{phi}-\eqref{constraint}, we use the Chebyshev pseudo spectral method along the $z$ direction and the Fourier pseudo spectral method in the $y$ direction. We find that along the $x$ direction, the pseudo spectral method that has been exclusively in holography cannot properly accommodate the interface dynamics. Instead, we choose the fourth order finite difference scheme along the $x$ direction in order to better capture the dynamics near the interface. For a typical simulation, we work on a $1001\times 101$ grid points along the two boundary spatial directions together with $20$ grid points in the holographic direction.

\section{Linear instability around a stationary configuration}\label{app:QNM}

We can examine how the stability of these stationary solutions starts to falter by using linear response theory. To be clear, we turn on small perturbations on the stationary background shown \emph{e.g.} in the left panel of Fig.~\ref{O_rho}.
\begin{equation}
\begin{aligned}
     \Phi_i=\Phi_{i0}+\delta\Phi_i,\quad A_\mu=A_{\mu 0}+\delta A_\mu\,, 
\end{aligned}
\end{equation}
where $\Phi_{0i}$ and $A_{\mu0}$ are stationary solutions.
Considering the fact that the background is the same over time and in the 
$y$-direction, we choose the bulk perturbation fields as:
    \begin{equation}
    \begin{aligned}
        &\delta\Phi_i=u_i(z,x)e^{-i(\omega t-ky)}e^{i(v_i)_yy}, \quad \delta\Phi_i^*=v_i(z,x)e^{-i(\omega t-ky)}e^{-i(v_i)_yy},\\&\delta A_t=a_t(z,x)e^{-i(\omega t-ky)},\quad\delta A_x=a_x(z,x)e^{-i(\omega t-ky)}, \quad\delta A_y=a_y(z,x)e^{-i(\omega t-ky)} ,
    \end{aligned}
    \end{equation}
where $(v_1)_y=(v_2)_y=v_y$. The resulting linear perturbation equations are given explicitly as

   \begin{equation}
    \begin{aligned}
        \label{pertubation}        &2iA_{t0}\partial_zu_i+2ia_t\partial_z\Phi_{0i}+i\partial_zA_{t0}u_i+i\partial_za_t\Phi_{0i}
        +\partial_z(f\partial_zu_i)-zu_i +\partial_x^2u_i-(k+(v_i)_y)^2u_i
        -i\partial_xA_{x0}u_i\\
        &-i\Phi_{0i}(\partial_xa_x+ika_y)-(A_{x0}^2+A_{y0}^2)u_i-2A_{x0}\Phi_{0i}a_x-2A_{y0}\Phi_{0i}a_y
        -2i(A_{x0}\partial_xu_i+i(k+(v_i)_y)A_{y0}u_i)\\&-2i(a_x\partial_x\Phi_{0i}+ia_y(v_i)_y\Phi_{0i})-\frac{\nu}{2}|\Phi_{0j}|^2u_i
        -\frac{\nu}{2} \Phi_{0j}^*\Phi_{0i}u_j-\frac{\nu}{2}  \Phi_{0j}\Phi_{0i}v_j\\&=-2i\omega\partial_zu_i, \qquad(i,j=1,2\quad i\ne j)\,,\\
        &-2iA_{t0}\partial_zv_i-2ia_t\partial_z\Phi_{0i}^*-i\partial_zA_{t0}v_i-i\partial_za_t\Phi_{0i}^*
        +\partial_z(f\partial_zv_i)-zv_i +\partial_x^2v_i-(k-(v_i)_y)^2v_i
        +i\partial_xA_{x0}v_i\\
        &+i\Phi_{0i}^*(\partial_xa_x+ika_y)-(A_{x0}^2+A_{y0}^2)v_i-2A_{x0}\Phi_{0i}^*a_x-2A_{y0}\Phi_{0i}^*a_y
        +2i(A_{x0}\partial_xv_i+i(k-(v_i)_y)A_{y0}v_i)\\&
        +2i(a_x\partial_x\Phi_{0i}^*-ia_y(v_i)_y\Phi_{0i}^*)-\frac{\nu}{2}|\Phi_{0j}|^2v_i
        -\frac{\nu}{2} \Phi_{0j}\Phi_{0i}^*v_j-\frac{\nu}{2}  \Phi_{0j}^*\Phi_{0i}^*u_j\\&=-2i\omega\partial_zv_i, \qquad(i,j=1,2\quad i\ne j)\,,\\
        &\partial_x^2a_t-k^2a_t+f\partial_z\partial_xa_x+ikf\partial_za_y-2a_t\sum_i|\Phi_{0i}|^2-2A_{t0}\sum_i(\Phi_{0i}^*u_i+\Phi_{0i}v_i)+if\sum_i(\Phi_{0i}^*\partial_zu_i\\
        &-\Phi_{0i}\partial_zv_i+v_i\partial_z\Phi_{0i}-u_i\partial_z\Phi_{0i}^*)=-i\omega(\partial_za_t+\partial_xa_x+ika_y)+\omega\sum_i(\Phi_{0i}^* u_i-\Phi_{0i} v_i)\,,\\
        &\partial_z(\partial_xa_t+f\partial_za_x)-(k^2a_x+ik\partial_xa_y)-2a_x\sum_i|\Phi_{0i}|^2-2A_{x0}\sum_i(\Phi_{0i}^*u_i+\Phi_{0i}v_i)\\
        &-i\sum_i(\Phi_{0i}^*\partial_xu_i-\Phi_{0i}\partial_xv_i+v_i\partial_x\Phi_{0i}-u_i\partial_x\Phi_{0i}^*)=-2i\omega\partial_za_x,\\
        &ik\partial_za_t+\partial_z(f\partial_za_y)+\partial_x^2a_y-ik\partial_xa_x-2a_y\sum_i|\Phi_{0i}|^2-2A_{y0}\sum_i(\Phi_{0i}^*u_i+\Phi_{0i}v_i)\\
        &+\sum_i((k+(v_i)_y)\Phi_{0i}^*u_i-(k-(v_i)_y)\Phi_{0i}v_i+(v_i)_yv_i\Phi_{0i}+(v_i)_yu_i\Phi_{0i}^*)=-2i\omega\partial_za_y\,.
    \end{aligned}
   \end{equation}

For improved numerical stability, we utilize the following equation for  $a_t$:
\begin{equation}
    \begin{aligned}
    &\partial_z(\partial_xa_x+ika_y-\partial_za_t)+i\sum_i(\Phi_i^*\partial_zu_i+v_i\partial_z\Phi_i-u_i\partial_z\Phi_i^*-\Phi_i\partial_zv_i)=0\,,
    \end{aligned}
\end{equation}
which is derived from the constraint equation~\eqref{constraint}. Additionally, we demand that the last perturbation equation of~\eqref{pertubation} holds at the AdS boundary $z=0$, resulting in
\begin{equation}
(\partial_z\partial_xa_x+ik\partial_za_y=-i\omega \partial_za_z)|_{z=0}\,. 
\end{equation}
Then, by considering~\eqref{relation}, the last perturbation equation is also satisfied throughout the bulk. Concerning other perturbed fields, we impose the source-free boundary condition at the AdS boundary. We further apply the Neumann boundary condition at $x=\pm L_x/2$. 

\begin{figure}[h]
        \centering
        \includegraphics[width=0.7\linewidth]{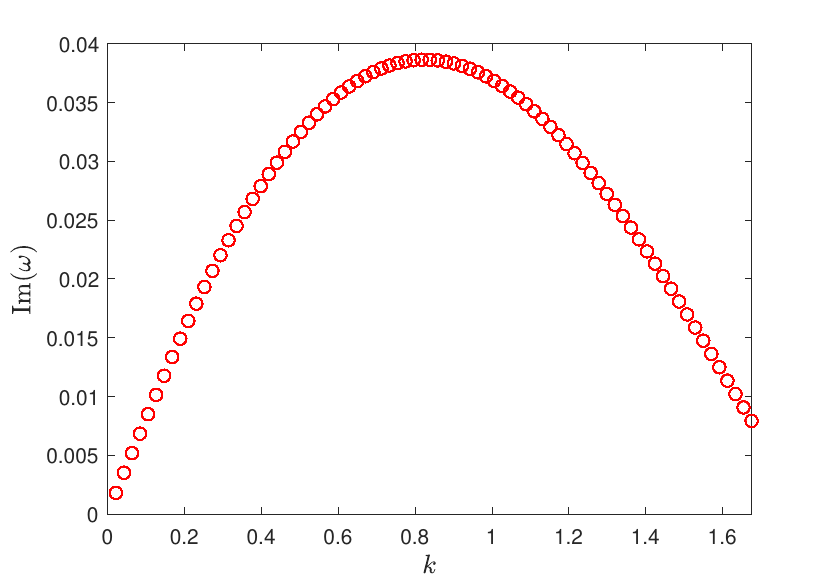}
        \caption{The QNMs spectrum for the stationary phase-separated configuration with $\mu=6$, $\nu=1$ and $v_y=1.2566$. $\mathrm{Im}(\omega)$ increases with the rise of $k$, peaking at a specific wave number corresponding to the fastest-growing mode.}
    \label{fig:omega}
\end{figure}

The corresponding QNMs are obtained by solving the above generalized eigenvalue problem. Then we can numerically determine $\omega$ for each $k$ and velocity $v_y$. Due to dissipations into the normal component, the quasi-normal frequencies generally take complex values. Note that the complex conjugates of linearized equations of motion can be obtained via the following transformation:
 \begin{equation}
     k\rightarrow -k,\quad \omega\rightarrow -\omega^*, \quad u_i\leftrightarrow v^*_i ,\quad a_\mu \rightarrow a_\mu^*\,,
 \end{equation}
 so that whenever $\omega$ is an eigenvalue for given wave number $k$, $-\omega^*$ is an eigenvalue for $-k$, and they share the same imaginary part. Therefore, without loss of generality, we only need to consider positive $k$. Since $\delta\Phi_i\sim e^{-i\omega t}$, the stationary configuration will become dynamically unstable whenever $\mathrm{Im}(\omega)>0$. The larger the positive imaginary part is, the more unstable the system becomes. In Fig.~\ref{fig:omega}, we show the spectrum of QNMs as a function of $k$ for $v_y=1.2566$ at $T/T_c=0.677$ and $\nu=1$. The imaginary part increases with $k$, peaks at a certain wave number that corresponds to the fastest growing mode. 

\section{On the origin of co-flow interface instability}\label{thermo}

The fact that imaginary part of QNMs is positive itself indicates a dynamic instability in the sense that the perturbation grows exponentially with time. Whether such dynamic instability corresponds to a thermal instability should be determined from an independent analysis of thermodynamics of the system.

Analysing thermodynamics of the binary superfluids would much improve our understanding of the system. 
As shown in Fig.~\ref{O_rho}, the system is highly inhomogeneous near the interface. Nevertheless, let's assume our system is in local thermal equilibrium and can still be properly described by thermodynamics as in the homogeneous case. Then, following the notation in~\cite{Gouteraux:2022qix}, the first law for binary superfluids reads 
\begin{equation}
    d\epsilon=Tds+\mu d\rho+\bm{\mathrm{v^n_1}}d\bm{\mathrm{g_1}}+\bm{\mathrm{h_1}}d\bm{\mathrm{v^s_1}}+\bm{\mathrm{v^n_2}}d\bm{\mathrm{g_2}}+\bm{\mathrm{h_2}}d\bm{\mathrm{v^s_2}}\,,
\end{equation}
where $\epsilon$, $s$, $\bm{\mathrm{g_i}}$ are the energy, entropy and momentum densities, and $\bm{\mathrm{h_i}}=\rho^s_i(\bm{\mathrm{v^s_i}}-\bm{\mathrm{v^n_i}})$. Here $\bm{\mathrm{v^s_i}}$ and $\bm{\mathrm{v^n_i}}$ are the velocities of superfluid and normal fluid components for the $i$-th superfluid. In our work, there is a further constraint $\bm{\mathrm{v^s_1}}=\bm{\mathrm{v^s_2}}$, which gives us $d\bm{\mathrm{v^s_1}}=d\bm{\mathrm{v^s_2}}$. Moreover, in the probe limit, one has $\bm{\mathrm{v^n_1}}=\bm{\mathrm{v^n_2}}=0$ in our setup. Let's denote $\bm{\mathrm{v^s_1}}=\bm{\mathrm{v^s_2}}=\bm{\mathrm{v^s}}$. Then the first law is modified to 
\begin{equation}
    d\epsilon=Tds+\mu d\rho+(\rho^s_1+\rho^s_2)\bm{\mathrm{v^s}}d\bm{\mathrm{v^s}}\,.
\end{equation}
With this in hand, one can see the thermodynamic susceptibilities are the same as those in~\cite{Gouteraux:2022qix}, except $\rho^s\rightarrow \rho^s_1+\rho^s_2$. Therefore the criterion
for thermodynamic instability would be 
\begin{equation}
\label{criterion}
    \partial_{\bm{\mathrm{v^s}}}[(\rho^s_1+\rho^s_2)\bm{\mathrm{v^s}}]|_{\bm{\mathrm{v^s}}=\bm{\mathrm{v^s}}_c}=0\,.
\end{equation}
Since in our model, two scalar fields are coupled to a single gauge field, from holographic dictionary, we can only read off the sum of the two components
\begin{equation}
\label{sum}
    \sum_i(\rho^s_iv^s_i)=\mu\partial_zA_y|_{z=0}/|v^s|\,.
\end{equation}
where we have considered that that the interface and superfluid velocities are along $y$ direction. In our co-flow binary superfluids, Eq.~\eqref{sum} gives exactly $(\rho^s_1+\rho^s_2)v^s$ which is needed to determine the critical velocity by using~\eqref{criterion}.

As can be seen from the left panel of Fig.~\ref{O_rho}, far from the interface, the system in thermal equilibrium is described by the homogeneous and isotropic hairy black brane solution with single condensate.
In this region, we can take directly the criterion $\partial_{\bm{\mathrm{v^s}}}(\rho^s_i\bm{\mathrm{v^s}})|_{\bm{\mathrm{v^s}}=\bm{\mathrm{v^s}}_c}=0$ which gives precisely the Landau critical velocity in our model. In contrast, for our immiscible superfluids, by calculating $v^s_c$ following~\eqref{criterion}, we find that it does not give the correct criterion for the onset of the interface instability.

\bibliographystyle{JHEP}
\bibliography{refs}

\end{document}